\documentclass[10pt,aps,prl,twocolumn,superscriptaddress,nofootinbib]{revtex4-1}
\usepackage{graphicx}
\usepackage{bigints}
\usepackage[normalem]{ulem}
\usepackage{cancel}
\usepackage{mathtools}
\usepackage{verbatim}
\usepackage{slashed}
\usepackage{titlesec}
\usepackage[usenames]{xcolor}
\usepackage{amsmath,amssymb}
\usepackage{mathrsfs}
\usepackage{amsthm}
\usepackage{hyperref}
\usepackage{appendix}
\usepackage{array}
\usepackage{booktabs,chemformula}

\newcommand{\p}{\partial}

\newcommand{\non}{\nonumber}

\newcommand{\be}[1]{ \begin{equation}\label{#1} }
\newcommand{\ee}{\end{equation}}
\newcommand{\bea}[1]{\begin{eqnarray}\label{#1} }
\newcommand{\eea}{\end{eqnarray}}
\newcommand{\bes}{\begin{subequations}}
\newcommand{\ees}{\end{subequations}}

\newcommand{\D}{\Delta}

\begin{document}

\title{Carrollian expansion of Born-Infeld Electrodynamics}
\author{Aditya Mehra}
\email{apm.sbas@jspmuni.ac.in}
\affiliation{School of Basic and Applied Sciences, JSPM University, Gate No. 720, Wagholi, Pune 412207,
India
}

\author{Hemant Rathi}
\email{hemant.rathi@saha.ac.in}
\affiliation{Saha Institute of Nuclear Physics, 1/AF Bidhannagar,
Kolkata 700064, India}

\author{Dibakar Roychowdhury}
\email{dibakar.roychowdhury@ph.iitr.ac.in}
\affiliation{Department of Physics, Indian Institute of Technology Roorkee, Roorkee 247667, Uttarakhand, India}

\begin{abstract}
In this paper, we carry out a detailed investigation of Born-Infeld Electrodynamics in the Carrollian limit. We explore these theories by looking at expansion in a small $c$ limit at
the level of Lagrangian. Particularly, we study the symmetries
and correlation functions in  the presence of non-linear interactions.
\end{abstract}	
\maketitle

\small

\section{Introduction}
Maxwell's Electrodynamics in four spacetime dimensions exhibits a remarkable property namely the conformal invariance along with the $U(1)$ gauge symmetry. On top of this, the corresponding equations of motion are linear in the field strength $F_{\mu\nu}$. From further studies, it has been confirmed that a generic quantum theory of Electrodynamics should contain higher order corrections (arising from the loop contributions) to the terms linear in $F_{\mu\nu}$ that is present in the original Lagrangian \cite{heisenberg2006consequences}. In the low energy limit, it boils down into the pure Maxwell term. This makes us wonder, if there exists other Lagrangian descriptions for $U(1)$ gauge fields which are non-linear in $F_{\mu\nu}$ and boils down to a linear theory as an effective description. These theories give rise to new physics at different energy scales. Below, we focus on one particular example that goes under the name of the Born-Infeld (BI) theory of Electrodynamics. 

Born-Infeld theory is a standard example of non-linear Electrodynamics\footnote{Another interesting example of
non-linear Electrodynamics is the ModMax Electrodynamics \cite{Bandos:2020jsw, Bandos:2021rqy, Avetisyan_2021, Rathi:2023vhw}. It is the non linear generalisation of Maxwell's Electrodynamics that preserves both $SO(2)$ and conformal invariance. }.  It was first introduced by Max Born and Leopold Infeld \cite{Born:1934gh, B2, M1, D1} to remove the divergences in the electron's self-energy in the Maxwell's theory of Electromagnetism by introducing an upper bound on the electric field strength at the origin. BI Electrodynamics can also be obtained from string theory in its low energy limit \cite{Tseytlin:1986ti, Fradkin:1985qd, Polchinski_1995} as an expansion in powers of derivatives of gauge fields. Finally, BI theory shows good physical properties such as the wave propagation as in the case of Maxwell's theory \cite{osti_4071071}.

In the present paper, we explore the Carrollian limit or the ultra-relativistic limit of BI Electrodynamics. Carrollian limit was established in \cite{Levy1965,1966ND} where we take the speed of light $(c)$ going to zero. This limit is a classic example of what is known as the ultra local Quantum Field Theories (QFT). The other extreme corner turns out to be the famous Galilean limit $(c \to \infty)$. In the language of the causal structure of the space-time, the Carrollian limit corresponds to the closure of the light cone along the time direction. A major effect of Carrollian limit on the Poincar\'{e} transformation is that it considers the space as absolute. Under this framework, the notion of causality completely goes away and the only possible way for two events to talk to each other is when they are at 
same space and time point. For this reason, Carroll limit is sometimes referred as the ``ultra-local'' limit of relativistic QFTs.

As said before, the Carrollian generators can be obtained from conformal symmetry generators following a $c \to 0$ contraction.
Taking such ultra-relativistic limit has attained renewed interest during past one decade, where there has been a significant amount of research activities in the area of ultra-local Quantum Field theories that possess an underlying (conformal) Carrollian symmetry as its relativistic conformal counterpart (see \cite{Bagchi:2019clu, Banerjee:2020qjj, Baiguera:2022lsw, Chen:2023pqf, Bagchi:2022eav}). Carrollian symmetry has made its way into many physical systems including black holes \cite{Donnay:2019jiz, Gray:2022svz}, plane gravitational waves \cite{Duval:2017els},  condensed matter \cite{Marsot:2022imf, Bagchi:2022eui} and also it has been portrayed as the candidate for flat space holography program \cite{Donnay:2022aba, Bagchi:2022emh, Bagchi:2016bcd, Duval:2014uva}. On the other hand, the modified versions of Born-Infeld theories have been hypothesised to be used in tachyon cosmology \cite{Gibbons_2002}. 

The purpose of the present article is to explore the Carrollian limit of BI Electrodynamics in the electric sector by taking the small $c$ expansion at the level of action \cite{deBoer:2021jej}. In particular, we study the symmetries and the correlation functions in the presence of non-linear interactions.

The organisation for the rest of the paper is as follows:
\begin{itemize}

\item In Section 2, we obtain the Carrollian limit of Born-Infeld (CBI) Electrodynamics in small $c$ expansion.
\item In Section 3, we investigate the symmetries of CBI Electrodynamics in the electric limit and compute the 2-point correlation between the components of $U(1)$ gauge fields. 
\item In Section 4, we discuss the symmetries of CBI theory at an
arbitrary order in the BI coupling.
\item Finally, in Section 5, we conclude along with some future directions.
\end{itemize}

\section{Carrollian limit of Born-Infeld Electrodynamics}

We explore the Carrollain limit of BI Electrodynamics in the electric limit in four-dimensional space-time in small $c$ expansion.

Born-Infeld Electrodynamics is a non-linear extension of the standard Maxwell Electrodynamics \cite{Born:1934gh, B2, M1, D1}. This theory aimed to cure the divergences in the electric field self-energy of the electron. In order to obtain the finite self-energy of a charged particle, the authors in \cite{Born:1934gh} put an upper bound on the electric field at the origin and derive the Lagrangian from the general invariance principle. The BI Lagrangain is given below
\bea{}\label{lagbi1} &&\mathcal{L}=-b^2 \sqrt{-\text{det}\Bigg( \eta_{\mu\nu}+\frac{1}{b}F_{\mu\nu}\Bigg)} +b^2,\eea
where $b$ is the BI coupling having the dimension of energy density and $F_{\mu\nu}=\partial_{\mu}A_{\nu}-\partial_{\nu}A_{\mu}$ is the standard electromagnetic field tensor.

Alternatively, one can also express the above Lagrangian density (\ref{lagbi1}) in the following form \cite{Zwiebach:2004tj} 
\bea{}\label{lagbi}&& \mathcal{L}= -b^2 \sqrt{1- \frac{2s}{b^2}-\frac{p^2}{b^4}} +b^2, \eea
where $s=-\frac{1}{4}F^{\mu\nu}F_{\mu\nu},~ p=-\frac{1}{4}\tilde{F}^{\mu\nu}F_{\mu\nu}$ along with $\tilde{F}^{\mu\nu}=\frac{1}{2}\epsilon^{\mu\nu\rho\sigma}F_{\rho\sigma}$. Notice that, in the large $b$ limit, the BI theory (\ref{lagbi}) reduce to the standard Maxwell Electrodynamics.

Using the Lagrangian density (\ref{lagbi}), one can estimate the upper bound on the electric field by demanding the expression inside the square root to be a positive quantity \cite{No1}. It turns out that the maximum allowed value of the electric field is of the order of BI coupling ($b$).

Later, it was found that the BI action emerges as an effective low-energy action in the open bosonic string theory \cite{Fradkin:1985qd}. Moreover, in the context of type II superstring theories, the BI action effectively describes the D-branes world volume theory in 10 space-time dimensions, where the BI coupling is related to the brane tension \cite{Leigh}.

We will now move on to look at the symmetries of BI action. It is noteworthy to mention that the BI Lagrangian (\ref{lagbi}) is invariant under the gauge and Poincare transformations
\bes{}\label{poinc}
\bea{}
&&\hspace{-1cm} {\delta}^{\mbox{\tiny{G}}} \, A_{\mu} (x) =  \p_\mu \alpha (x),~~
{\delta}_\mu^{\mbox{\tiny{P}}} \, A_{\nu} (x) = \p_\mu A_{\nu} (x), \\ &&\hspace{-1cm}{\delta}_{\mu \nu}^{\mbox{\tiny{L}}} \, A_{\rho} (x) = (x_\mu \p_\nu - x_\nu \p_\mu)A_{\rho}(x) + \eta_{\rho \mu} A_{\nu}(x)-\eta_{\rho \nu} A_{\mu}(x).
\eea
\ees
However, if we try to look into the invariance of \eqref{lagbi} under the conformal extension of Poincare algebra that includes scale $(D)$ and special conformal transformations $(K_{\mu})$.  The transformations are given by
\bes{}\label{poinc1}
\bea{}&&{\delta}^{\mbox{\tiny{D}}} \, A_{\mu} (x) =(x^{\nu}\p_{\nu}+\D)A_{\mu},\\&&\non
{\delta}_\mu^{\mbox{\tiny{K}}}A_{\nu}(x)=(2x_{\mu}x^{\tau}\p_{\tau}-x^2\p_{\mu}+2\D x_{\mu})A_{\nu}\\&&\hspace{2.5cm}-2(x_{\nu}A_{\mu}-\eta_{\mu\nu}x^{\tau}A_{\tau}),
 \eea\ees
where $\D$ is the scaling weight of the fields under consideration. It is not invariant under the scale and the special conformal transformations. It concludes that the relativistic Born-Infeld theory is only invariant under Poincare transformations and not under its conformal extension.

Next, we study the small $c$ expansion at the Lagrangian level. In other words, we first have to expand the relativistic Lagrangian in $c$ and then take the $c\rightarrow 0$ limit in order to get the Carrollian version. Moreover, it is technically challenged to investigate the symmetries of CBI Electrodynamics for a general value of BI coupling due to the presence of square root in the Lagrangian density (\ref{lagbi}). Therefore, we will study the symmetries of the theory perturbatively in the BI coupling (equivalently, in large $b$ limit). In the large $b$ limit, the above Lagrangian density (\ref{lagbi}) boils down\footnote{For higher order corrections, see Section \ref{HOC} }
\begin{align}\label{la1}
    \mathcal{L}=s+\frac{1}{2b^2}\left(s^2+p^2\right)+O(1/b^4),
\end{align}
where the explicit form of $s$ and $p$ are given as 
\begin{align}
    s&=\frac{1}{2c^2}\left(\partial_t A_i-\partial_iA_t\right)^2-\frac{1}{4}\left(\partial_i A_j-\partial_jA_i\right)^2,\\
    p&=-\frac{1}{2c}\left(\partial_tA_i-\partial_iA_t\right)\left(\partial_j A_k-\partial_kA_j\right)\epsilon^{0ijk},
\end{align}
where we identify $\partial_0=\frac{1}{c}\frac{\partial}{\partial t}$, $\partial_i=\frac{\partial}{\partial x^{i}}$ and $A_t=cA_0$.

Now we make an expansion of the Lagrangian density (\ref{la1}) around $c=0$. The components of gauge fields can be expanded around $c=0$ as follows 
\begin{align}
    A_t&=c^{\Delta}\left(A_t^{(0)}+c^2A_{t}^{(1)}+...\right),\label{laA1}\\
A_i&=c^{\Delta}\left(A_i^{(0)}+c^2A_{i}^{(1)}+...\right),\label{laA2}
\end{align}
where $\Delta$ is the scaling dimension.

Using (\ref{laA1}) and (\ref{laA2}) in (\ref{la1}), the leading order Lagrangian can be systematically expressed as 
\begin{align}
    \mathcal{L}=c^{2\Delta-2} \mathcal{L}_{\text{Maxwell}}+ c^{4\Delta-4}\mathcal{L}_{\text{Born Infeld}},\label{lalag}
\end{align}
where we define
\begin{align}
\mathcal{L}_{\text{Maxwell}}=\mathcal{L}_{\text{Maxwell}}^{\text{(E)}}+\mathcal{L}_{\text{Maxwell}}^{\text{(M)}}\hspace{1mm},\hspace{2mm} \mathcal{L}_{\text{BI}}=\mathcal{L}_{\text{BI}}^{\text{(E)}}+\mathcal{L}_{\text{BI}}^{\text{(M)}},\label{LP00}
\end{align}
where
\bea{}&&
    \mathcal{L}_{\text{Maxwell}}^{(E)}=\hspace{1mm}\frac{1}{2}\left(E_i^{(0)}\right)^2\hspace{1mm},\hspace{2mm}\mathcal{L}_{\text{BI}}^{(E)}=\frac{1}{8b^2}\left(E_i^{(0)}\right)^4,\label{LP01}\\&&
\mathcal{L}_{\text{Maxwell}}^{(M)}=\hspace{1mm}c^2\left(E_i^{(0)}E_i^{(1)}-\frac{1}{4}\left(F_{ij}^{(0)}\right)^2\right), \label{LP02}\\&&\non
\mathcal{L}_{\text{BI}}^{(M)}=\hspace{1mm}\frac{c^2}{8b^2}\Bigg[\left(E_j^{(0)}\right)^2\left(4E_i^{(0)}E_i^{(1)}-\left(F_{ij}^{(0)}\right)^2\right)\\&&\hspace{3.8cm}+\left(E_i^{(0)}F_{jk}^{(0)}\epsilon^{0ijk}\right)^2\Bigg].\label{LP03}
\eea
Here, where we define 
\begin{align}
    E_i^{(n)}=\partial_i A_{t}^{(n)}-\partial_{t}A_i^{(n)},\hspace{2mm}
    F_{ij}^{(n)}=\partial_i A_{j}^{(n)}-\partial_{j}A_i^{(n)},
\end{align}
and $n$ can take values either 0 or 1. We call this Lagrangian as Carrollian Born Infeld (CBI) Electrodynamics. Furthermore, the superscripts $(E)$ and $(M)$ in $\mathcal{L}_{\text{Maxwell}}$ and $\mathcal{L}_{\text{BI}}$ (\ref{LP00}) respectively denotes the Lagrangian in the electric and magnetic limit. It should be noted that the electric limit appears at the leading term in (\ref{LP00}), where the magnetic effects start appearing at the sub-leading term (coefficient of $c^2$). For the present purpose, we focus only on the electric limit.

The equation of motions in the electric limit can be obtained by varying the Lagrangian density (\ref{lalag}) with respect to the gauge fields ($A_t^{(0)},A_i^{(0)}$) as 
\begin{align}
    \partial_i\left(E_i^{(0)}+\frac{1}{2b^2}E_i^{(0)}\left(E_j^{(0)}\right)^2\right)&=0,\label{lae1}\\
    \partial_t\left(E_i^{(0)}+\frac{1}{2b^2}E_i^{(0)}\left(E_j^{(0)}\right)^2\right)&=0.\label{lae2}
\end{align}

\section{Symmetries and Correlation functions}

In this Section, we discuss the symmetries of the CBI Electrodynamics in the electric limit and using the symmetries of the theory, we will compute the correlation functions between the components of $ U(1)$ gauge fields. 

\subsection{Symmetries}

In this Section, we discuss the invariance of the CBI action (\ref{LP00}) under the Carrollian generators and its conformal extensions. 

The Carrollian generators includes translations $(P_i,H)$, rotations $(J_{ij})$, boosts $(B_i)$ and its conformal extensions contains scale transformation $(D)$, special conformal transformations $(K,K_i)$. Together they form what we call as finite Conformal Carrollian algebra (fCCA). It is also possible to give a lift to the finite CCA. In place of fifteen generators, we will now have infinite number of generators and still the CCA will close among itself. We call it as infinite dimensional CCA. For this paper, we are interested in $d=4$ case only. The infinite extension was proposed only in the supertranslation part \cite{Bagchi_2016}:
\bea{}
M^{m_1,m_2,m_3} =x^{m_1}y^{m_2}z^{m_3}\p_t, 
\eea
where $m_1,m_2,m_3$ are integers. These generators correspond to the time translation generators with arbitrary spatial dependencies. The finite parts of supertranslations are given by:
\bes{}
\bea{}&&\hspace{-1cm}
 H=M^{0,0,0},~
B_x=M^{1,0,0}, \ B_y=M^{0,1,0}, \ B_z=M^{0,0,1}, \\&&\hspace{-1cm} K=M^{2,0,0}+M^{0,2,0}+M^{0,0,2}. 
\eea \ees
The infinite dimensional CCA (I-CCA) will include the finite part along with the supertranslations $(M)$. The Carrollian transformations on the gauge fields can be found by taking the Carrollian limit directly on \eqref{poinc} and \eqref{poinc1}.

The CBI action (\ref{LP00})  is completely invariant under translations $(\delta_{P}\equiv \p_i, \delta_{H}\equiv \p_t)$ and rotations $(\delta_J \equiv (x_i\p_j -x_j \p_i +S_{ij})$, where $S_{ij}$ is the spin matrix. Under the Carrollian boost, the components of gauge field transform as \cite{deBoer:2021jej} 
\bes{}
\bea{}\hspace{-.30cm}
    \delta A_t^{(n)}&=\Vec{g}.\Vec{x}\partial_tA_{t}^{(n)}+t\Vec{g}.\partial_tA_{t}^{(n-1)}+\Vec{g}.A_{t}^{(n-1)},\label{lasb1}\\
     \delta A_i^{(n)}&=\Vec{g}.\Vec{x}\partial_tA_{i}^{(n)}+g_iA_{t}^{(n)}+t\Vec{g}.\partial A_{t}^{(n-1)},\label{lasb2}
\eea\ees
where $\Vec{g}$ is the Carroll boost parameter and $n$ can take values either 0 or 1.

Using (\ref{lasb1}) and (\ref{lasb2}), it is now quite straight forward to check that the Lagrangian (\ref{LP00}) in the electric limit is invariant under the Carrollian boost (up to total derivative). Moreover, the equations of motion (\ref{lae1})-(\ref{lae2}) also remains invariant under (\ref{lasb1})-(\ref{lasb2}).

Next, we explore the variation of (\ref{LP00}) under the scale transformation ($D$). The components of gauge field ($A_{i},A_{t}$) and the field strength under dilatation ($D$) are given by 
\bea{}\label{dilatation}
   && \delta_{D}A_K=(t\partial_t+x_k\partial_k+\Delta)A_K,
\eea
where $K=t,i$.
Notice that, the electric piece of the Lagrangian (\ref{LP00}) is not invariant under the scale transformation (\ref{dilatation}), rather it turns out to be
\begin{align}\label{sm01}
\delta_D\left(\mathcal{L}_{\text{Maxwell}}^{(E)}+\mathcal{L}_{\text{BI}}^{(E)}\right)=\frac{1}{2b^2}\left(E_i^{(0)}\right)^4,
\end{align}
which vanishes in the strict $b\rightarrow\infty$ limit. This is what we expect from the standard Carrollian Electrodynamics \cite{deBoer:2021jej}. 

Under the temporal component of special conformal transformation $(K)$ and Carrollian supertranslations $(M^{m_1,m_2,m_3})$ given by 
\bes{}
\bea{}&&\hspace{-1cm}\delta_{M} A_t= x^{m_1}y^{m_2}z^{m_3}\p_t A_t,\\&&\hspace{-1cm} \delta_{M} A_l= x^{m_1}y^{m_2}z^{m_3}\p_t A_l+\partial_l\big(x^{m_1}y^{m_2}z^{m_3}\big) A_t,\eea\ees
the action (\ref{LP00}) remains invariant. On the other hand, it is not invariant under the spatial component of special conformal generators ($K_j$):
\bes{}
\bea{}&&\hspace{-1cm}\delta_{K_j} A_t= (2\D x_j+2x_jt\partial_t+2x_i x_j \partial_i- x_i x_i \partial_j ) A_t,\\&&\hspace{-1cm}\non
\delta_{K_j} A_l= (2\D x_j+2x_jt\partial_t+2x_i x_j \partial_i- x_i x_i \partial_j )A_l\\&&\hspace{1.5cm}+2\delta_{jl} x^i A_i-2x_l A_j+2t \delta_{lj} A_t.\eea\ees

\subsection{Correlation functions}
In this Section, we compute the 2-point correlation between the different components ($A_t,\hspace{1mm} A_i$) of the $U(1)$ gauge field in the electric limit. In order to compute these correlation functions, we use the symmetry generators in the Carrollian BI limit that preserve the structure of the action up to leading order (LO) in the BI coupling.

Before proceeding further, it is customary to discuss the algorithm to compute the 2-point correlation function based on \cite{Banerjee:2020qjj}. Firstly, we demand that the vacuum state $(|0\rangle)$ be invariant under all symmetry operators ($O$). Here, $O$ collectively denotes the temporal and spatial translations ($H,P_i$), rotation ($J_{ij}$), boost ($B_i$), temporal special conformal transformation ($K$) and Carrollian supertranslations ($M^{m_1m_2m_3}$).

The invariance of the vacuum state under symmetry operators ($O$) implies that $O|0\rangle=0$ and $\langle 0|O=0$. Using this fact, one can write down the master equation for computing the 2-point correlation \cite{Banerjee:2020qjj}
\begin{align}\label{cf1}
    \langle0|[O,f(t_1,x_1)]g(t_2,x_2)|0\rangle+\langle0| f(t_1,x_1)[O,g(t_2,x_2)]|0\rangle=0,
\end{align}
where $f(t,x)$ and $g(t,x)$ represent the components of gauge field ($A_t,\hspace{1mm} A_i$). Now, one can easily solve the above differential equations (\ref{cf1}) for various symmetry generators ($O$) and obtain the required correlation functions. 

We now move on to estimate all the possible correlation functions in the Carrollian Born-Infeld limit. \\

$\bullet$ \underline{Case I : Correlation between $A_t(x_1,t_1)$ and $A_t(x_2,t_2)$}

The 2-point correlation between the temporal components of the gauge field is defined as
\begin{align}\label{cff1}
    G_{00}^{(E)}(t_1,t_2,x_1,x_2)\equiv\langle A_t(t_1,x_1)A_t(t_2,x_2)\rangle.
\end{align}
Notice that, the invariance under temporal and spatial translation implies that $G_{00}^{(E)}$ can only depend on the difference between the coordinates i.e. $\tau=t_1-t_2$ and $r^i=x^i_1-x^i_2$. Using this fact and using the boost operator in equation (\ref{cf1}), one finds
\begin{align}\label{cfe1}
    r_i\partial_{\tau}G_{00}^{(E)}=0.
\end{align}

Equation (\ref{cfe1}) suggests that $G_{00}$ can at most depend on $r_i$. However, one can further refine the form of $G_{00}^{(E)}$ by requiring that the correlation functions must be invariant under rotation, which yields
\begin{align}\label{cfe2}
    G_{00}^{(E)}=c_0 r^2,
\end{align}
where $c_0$ is an arbitrary constant and $r^2=r_1^2+r_2^2+r_3^2$. 

Finally, it is straightforward to check that the other symmetry operators i.e. temporal special conformal transformation ($K$) and Carrollian supertranslations ($M^{m_1m_2m_3}$) also produce the same result.\\

$\bullet$ \underline{Case II : Correlation between $A_t(x_1,t_1)$ and $A_i(x_2,t_2)$}

Two point correlation between the temporal and the spatial component of the gauge field can be expressed as
\begin{align}\label{cff2}
    G_{0i}^{(E)}(t_1,t_2,x_1,x_2)\equiv\langle A_t(t_1,x_1)A_i(t_2,x_2)\rangle.
\end{align}

Similar to the previous case, the temporal and spatial translation invariance demand that $G_{0i}^{(E)}$ can only depend on $\tau$ and $r^i$. Furthermore, the boost symmetry implies that
\begin{align}\label{cfe3}
    r_l\partial_\tau G_{0i}^{(E)}+\delta_{li}G_{00}^{(E)}=0.
\end{align}

Now we solve the above equation (\ref{cfe3}) for different values of $l$ and $i$. Notice that, for $l=1$ and $i=2$, we find $G_{02}^{(E)}=\mathcal{F}_1(r^i)$. However, for $l=i=2$, we obtain
\begin{align}
    G_{02}^{(E)}=-c_0\frac{r^2}{r_2}\tau+\mathcal{F}_2(r^i).
\end{align}

In order to obtain a unique solution $\Big(G_{02}^{(E)}\Big)$, we must therefore set $c_0=0$ together with $\mathcal{F}_1(r^i)=\mathcal{F}_2(r^i)=\mathcal{F}(r^i)$. Setting, $c_0=0$ implies that the correlation between the temporal components of gauge field $\Big(G_{00}^{(E)}\Big)$ must vanish in the electric limit.

The rotational symmetry further restricts the form of $G_{0i}^{(E)}$ as follows
\begin{align}
    G_{0i}^{(E)}=c_{i}r^2,
\end{align}
where $c_{i}$, $i=(1,2,3)$ are the constants. It is now straightforward to check that the temporal special conformal transformation ($K$) and the Carrollian supertranslations ($M^{m_1m_2m_3}$) also lead to the same conclusion. \\

$\bullet$ \underline{Case III : Correlation between $A_i(x_1,t_1)$ and $A_j(x_2,t_2)$}

The two point correlation between the spatial components of gauge field is defined as as
\begin{align}\label{cff3}
    G_{ij}^{(E)}(t_1,t_2,x_1,x_2)\equiv\langle A_i(t_1,x_1)A_j(t_2,x_2)\rangle.
\end{align}

Like before, the temporal and spatial translation invariance restrict the dependencies of $G_{ij}^{(E)}$ only to $\tau$ and $r^i$. Moreover, the invariance of $G_{ij}^{(E)}$ under the boost symmetry requires that 
\begin{align}\label{cfe4}
    r_l\partial_{\tau}G_{ij}^{(E)}+\delta_{li}G_{0j}^{(E)}+\delta_{lj}G_{i0}^{(E)}=0.
\end{align}

Next, we solve the above equation (\ref{cfe4}) for different values of $l,i$ and $j$. First, we consider a case with $l=1,i=2,j=3$. For, these values of ($l,i,j$), we find $G_{23}=\mathcal{K}_1(r^i)$. On the other hand, for the choice $l=i=2,j=3$, we obtain
\begin{align}
     G_{23}^{(E)}=-c_3\frac{r^2}{r_2}\tau+\mathcal{K}_2(r^i).
\end{align}

In order to obtain a unique solution of the correlation function $G_{23}$, we must therefore identify the function $\mathcal{K}_1(r^i)$ with $\mathcal{K}_2(r^i)$ and set, $c_3=0$. Similarly, one can check that uniqueness of the other correlation function i.e. $G_{12}^{(E)}$ and $G_{31}^{(E)}$ requires to set $c_2=c_1=0$. In other words, the two point correlation between the temporal and spatial components of the gauge field must vanish in the electric limit. 

Finally, the rotational symmetry refines the form of the correlation function $G_{ij}^{(E)}$ as
\begin{align}\label{ecnc1}
    G_{ij}^{(E)}=c\delta_{ij}r^2,
\end{align}
where $c$ is an arbitrary constant. One can also reach the same conclusion (\ref{ecnc1}) using the temporal special conformal transformation ($K$) and Carrollian supertranslations ($M^{m_1m_2m_3}$).

Notice that the correlations between the components of the $U(1)$ gauge field grow as $G_{ij}\sim r^2$, where $r$ measures the spatial separation between two points. On the other hand, in the conformal Carrollian Electrodynamics, it is inversely proportional to the distance of separation i.e. $G_{ij}\sim r^{-2}$, which reflects the scale invariance of the theory in the Carrollian limit. Moreover, the theory is ultra-local therefore $r$ cannot be arbitrarily large.

\section{Comments on higher order corrections}\label{HOC}

In this Section, we discuss the symmetries of the CBI Electrodynamics (in electric limit) at $n^{th}$ order in the BI coupling $\big(1/b^2\big)$. To begin with, we expand the Lagrangian density (\ref{lagbi}) in the powers of $\big(1/b^2\big)$ 
\begin{align}\label{la1h}
    \mathcal{L}=s+\frac{1}{2b^2}\left(s^2+p^2\right)+\frac{s}{2b^4}\left(s^2+p^2\right)+ ...
\end{align}

Next, we expand the above Lagrangian density (\ref{la1h}) around $c=0$ using the gauge fields expansions (\ref{laA1}) and (\ref{laA2}), which yields 
\begin{align}\label{higherl}
    \mathcal{L}^{(E)}=\sum_{n=0}^{\infty}\frac{c_n}{b^{2n}}\left(E_i^{(0)}\right)^{2n+2},
\end{align}
where $c_n$ are the constants.

The equation of motions associated with the above Lagrangian density (\ref{higherl}) is given below   
\begin{align}\label{quad1}
    \partial_{\alpha}\Bigg[E_i^{(0)}\sum_{n=0}^{\infty}\frac{c_n(2n+2)}{b^{2n}}\left(E_j^{(0)}\right)^{2n}\Bigg]=0,
\end{align}
where $\alpha=t,i$.

It is now quite straightforward to check that the action (\ref{higherl}) as well as the equations of motion (\ref{quad1}) remain invariant under the $K$, $B_l$, and $M$ transformations up to total derivative. For example, under the action of boost $B_l$, the above equation (\ref{quad1}) transforms as 
\begin{align}
    \partial_{\alpha}\Bigg[x_l\partial_t\Bigg\{E_i^{(0)}\sum_{n=0}^{\infty}\frac{c_n(2n+2)}{b^{2n}}\left(E_j^{(0)}\right)^{2n}\Bigg\}\Bigg],
\end{align}
which vanishes identically. Therefore, we conclude that the CBI Electrodynamics in the electric limit is invariant under $K$, $B_l$, and $M$ transformations for all orders in the BI coupling. Similarly, one can also explore the symmetries of the CBI Electrodynamics in magnetic limit. We leave this exciting direction for future investigations.

\section{Conclusion and future direction} \label{ConcS}
We now summarise the key findings of our analysis and outline some exciting projects down the line. 
We explore small $c$ expansion at the level of the Lagrangian \cite{deBoer:2021jej}. We then look into the Carrollian symmetries and find that the electric limit of Born-Infeld Electrodynamics is completely invariant. The surprising result is that the theory is invariant under supertranslations $(M^{m_1,m_2,m_3})$. Finally, we see how the correlation functions looks in the Carrollian limit for Born-Infeld Electrodynamics.

Below, we outline some further directions that might be worth pursuing in the future.

$\bullet$ It would be nice to explore the ``similar feature'' in the Carrollian limit of supersymmetric version of Born-Infeld theory and in particular, one can investigate the connection of this theory with tensionless limit \cite{Bagchi:2013bga, Bagchi:2015nca, Bagchi:2016yyf, Bagchi:2021rfw, Bagchi:2020fpr, Bagchi:2019cay} of string theory. 

$\bullet$ It would be an interesting project to study the ``similar feature'' in the Carrollian gravity coupled with the BI theory. In particular, the authors in \cite{deBoer:2023fnj} consider a Carroll gravity coupled with Maxwell Electrodynamics and construct various solutions, including the wormholes, black hole solutions etc., and examine the geodesic structure in the Carrollian limit. Therefore, it would be nice to explore the possible deviations in the presence of BI interactions in the theory.

\section*{Acknowledgments}
HR would like to thank the authorities of Saha Institute of Nuclear Physics, Kolkata, for their support. HR and DR are indebted to the authorities of Indian Institute of Technology, Roorkee for their unconditional support towards researches in basic sciences. DR would like to acknowledge The Royal Society, UK for financial assistance. DR would also like to acknowledge the Mathematical Research Impact Centric Support (MATRICS) grant (MTR/2023/000005) received from ANRF, India. AM wants to thanks JSPM university for their support during the progress of the project. AM also wants to thanks Arjun Bagchi and Poulami Nandi for initial collaboration and contribution in the project.

\bibliographystyle{unsrt}
\bibliography{Ref}

\end{document}